\documentclass[openacc]{rstransa}
\usepackage{graphicx, amsmath, amssymb, nicefrac, hyperref}
\usepackage[normalem]{ulem}
\usepackage{amsmath,siunitx}
\usepackage{caption}
\usepackage{subcaption}

\newcommand{\SiI}{Si~{\sc{i}}~10827{\,}{\AA}}

\begin{document}

\title{Magnetoacoustic Wave Energy Dissipation in the Atmosphere of Solar Pores}

\author{
Caitlin A. Gilchrist-Millar$^{1}$, David B. Jess$^{1,2}$, Samuel D. T. Grant$^{1}$, Peter H. Keys$^{1}$, Christian Beck$^{3}$, Shahin Jafarzadeh$^{4,5}$, \\ Julia M. Riedl$^{6}$, Tom Van Doorsselaere$^{6}$, Basilio Ruiz Cobo$^{7,8}$}

\address{$^{1}$Astrophysics Research Centre, School of Mathematics and Physics, Queen's University Belfast, Belfast, BT7 1NN, U.K. \\
$^{2}$Department of Physics and Astronomy, California State University Northridge, Northridge, CA 91330, U.S.A. \\
$^{3}$National Solar Observatory (NSO), 3665 Discovery Drive, Boulder, CO 80303, U.S.A. \\
$^{4}$Rosseland Centre for Solar Physics, University of Oslo, P.O. Box 1029 Blindern, N-0315 Oslo, Norway \\
$^{5}$Institute of Theoretical Astrophysics, University of Oslo, P.O. Box 1029 Blindern, NO-0315 Oslo, Norway \\
$^{6}$Centre for mathematical Plasma Astrophysics (CmPA), KU Leuven, Celestijnenlaan 200B bus 2400, 3001, Leuven, Belgium \\
$^{7}$Instituto de Astrof{\'{i}}sica de Canarias, V{\'{i}}a L{\'{a}}ctea s/n, 38205, La Laguna, Tenerife, Spain \\
$^{8}$Departamento de Astrof{\'{i}}sica, Universidad de La Laguna, 38206, La Laguna, Tenerife, Spain}

\subject{astrophysics, observational astronomy, solar system, spectroscopy, wave motion}


\corres{Caitlin A. Gilchrist-Millar\\
\email{cgilchristmillar01@qub.ac.uk}}

\begin{abstract}
The suitability of solar pores as magnetic wave guides has been a key topic of discussion in recent years. Here we present observational evidence of propagating magnetohydrodynamic wave activity in a group of five photospheric solar pores. Employing data obtained by the Facility Infrared Spectropolarimeter at the Dunn Solar Telescope, oscillations with periods on the order of 5~minutes were detected at varying atmospheric heights by examining {\SiI} line bisector velocities. Spectropolarimetric inversions, coupled with the spatially resolved root mean square bisector velocities, allowed the wave energy fluxes to be estimated as a function of atmospheric height for each pore. We find propagating magnetoacoustic sausage mode waves with energy fluxes on the order of 30~kW/m$^{2}$ at an atmospheric height of 100~km, dropping to approximately 2~kW/m$^{2}$ at an atmospheric height of around 500~km. The cross-sectional structuring of the energy fluxes reveals the presence of both body- and surface-mode sausage waves. Examination of the energy flux decay with atmospheric height provides an estimate of the damping length, found to have an average value across all 5 pores of $L_d \approx 268$~km, similar to the photospheric density scale height. We find the damping lengths are longer for body mode waves, suggesting that surface mode sausage oscillations are able to more readily dissipate their embedded wave energies. This work verifies the suitability of solar pores to act as efficient conduits when guiding magnetoacoustic wave energy upwards into the outer solar atmosphere. 
\end{abstract}



\maketitle

\section{Introduction}

The composition of solar pores, which are represented by strong magnetic field concentrations in the solar photosphere \cite{Simon1970}, make them ideal conduits for transporting wave energy from the lower solar atmosphere upwards into higher regions \cite{Jess2015}. When compared to sunspots, wave activity in these structures is further enhanced by their small size that makes them more susceptible to external forces, which promotes more dynamic wave generation \cite{Sobotka2003}. Numerical simulations have previously predicted that strong downflows (driven by cooling inside the pore centre) collide with the dense lower layers beneath the photosphere and rebound into upwardly propagating magnetoacoustic waves that can transport energy flux into the chromosphere \cite{Steiner1998, Cameron2007, Kato2011}. Such predictions have since been supported by observational evidence provided by both ground- and space-based observatories \cite{Cho2010, Cho2013}.  This highlights the importance of studying the photospheric counterpart of wave phenomena in order to be able to connect it with dynamics occurring in the outer layers of the solar atmosphere.

Oscillations manifesting in solar pores have been observed many times in the past. Early work, combining the Transition Region And Coronal Explorer (TRACE) satellite and German Vacuum Tower Telescope (VTT) observations, revealed oscillations of the photospheric magnetic field with periods spanning 5 -- 8 minutes, and corresponding chromospheric intensity oscillations with a period of approximately 3 minutes, suggesting the presence of upwardly propagating magnetoacoustic waves \cite{Balthasar2000}. 
In the case of magnetohydrodynamic (MHD) sausage mode waves, a characteristic feature is a periodic fluctuation in the cross-sectional area of the pore. Using an 11-hour time series of white light observations of a large pore, Dorotovi\u{c} et al. \cite{Dorotovic2002, Dorotovic2008} revealed periods in cross-sectional area in the range of 20 -- 70 minutes, providing evidence of slow sausage oscillations, potentially with elusive gravity waves acting as the main driver.

Simultaneous observations of fluctuations in both the intensity and cross-sectional area of pores was presented by Morton et al. \cite{Morton2011}, who performed a detailed study of photospheric sausage modes present in Rapid Oscillations in the Solar Atmosphere (ROSA) 4170{\,}{\AA} continuum observations of solar pores. In contrast to Dorotovic et al. \cite{Dorotovic2002, Dorotovic2008}, Morton et al. \cite{Morton2011} found much shorter periods in the range of 3 -- 5 minutes, indicating that they may be driven by a global pressure mode ($p$-mode) oscillation, rather than gravity mode waves. However, no conclusions were made about the specific wave mode (fast or slow) present, or whether these were standing or propagating waves. In a later study by Moreels et al. \cite{Moreels2013}, the same dataset studied by Morton et al. \cite{Morton2011} was re-examined, with the waves identified as fast modes due to the cross-sectional area and intensity variations being in antiphase to each other. On the contrary, in-phase cross-sectional area and intensity variations indicate the presence of a slow mode sausage wave. In subsequent work, Dorotovic et al. \cite{Dorotovic2014} detected both fast and slow sausage waves in three separate magnetic waveguides (two sunspots and one pore) in the deep photosphere, and proposed that the oscillations may be standing harmonics. Evidence of standing harmonics in a solar pore was also seen by Moreels et al. \cite{Moreels2015} through the application of MHD theory. Unfortunately, the simultaneous detection of the chromospheric counterparts to these photospheric waves was lacking. However, the general ubiquity of sausage mode waves in the chromosphere was demonstrated by Morton et al. \cite{Morton2012}, supporting the idea that MHD phenomena guided by magnetic pores may be able to contribute to upper atmospheric heating.

\begin{figure}[!t]
\centering\includegraphics[width=0.75\textwidth]{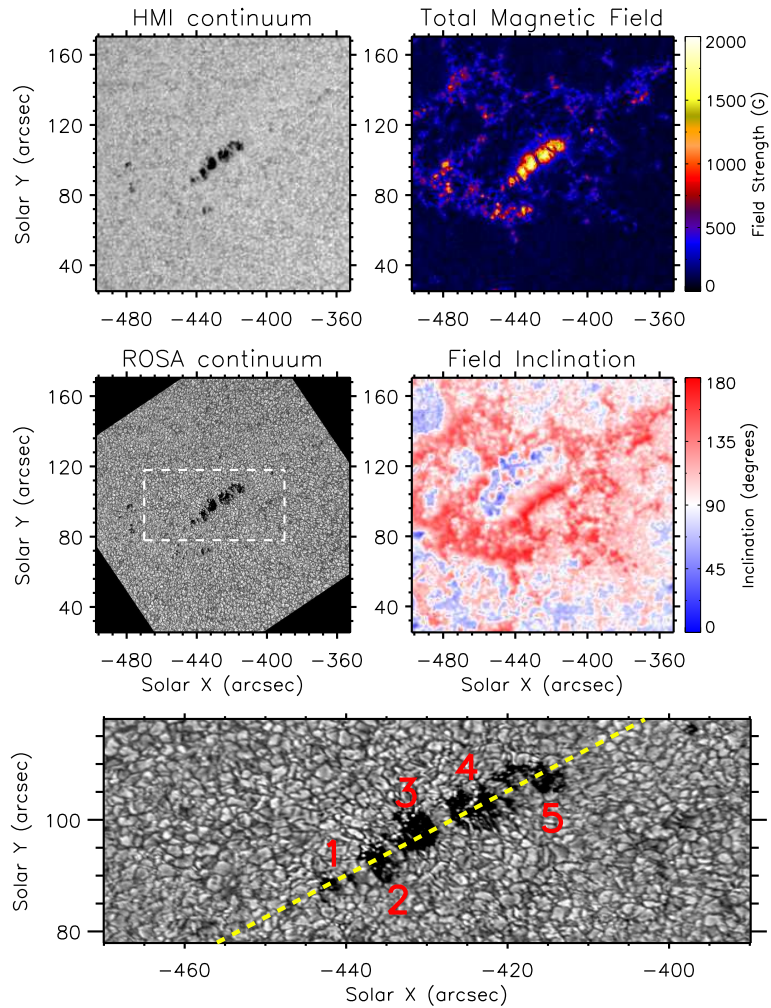}
\caption{HMI/SDO continuum image of active region NOAA~12564 (upper left), alongside the coaligned ROSA 4170{\,}{\AA} continuum image (middle left). Vector magnetogram products from HMI/SDO, including the total magnetic field strength and inclination angle to the solar normal are shown in the upper- and middle-right panels, respectively. An enlarged ROSA 4170{\,}{\AA} continuum image, obtained from within the white dashed lines shown in the middle-left panel, is displayed in the bottom panel. Here, the position of the FIRS slit is depicted by a dashed yellow line, with each of the 5 solar pores captured labelled by a number 1--5. The axes employed are all solar heliocentric coordinates.}
\label{fig_FOV}
\end{figure}

Combining multiple instruments and wavelength bands, Grant et al. \cite{Grant2015} were able to observe sausage mode oscillations, contained within the confines of a solar pore, propagating upwards from the lower photosphere to the base of the transition region. The authors utilised wavelet and Fourier techniques to detect oscillations in both the intensity and cross-sectional area of the pore, and observed periods in the range of 3 -- 7~minutes, remaining consistent with a global $p$-mode driver suggested by Morton et al. \cite{Morton2011}. The energy flux was calculated \cite{Moreels15b} to drop rapidly with height from the solar surface, with the initial energy flux of $\sim$35~kW{\,}m$^{-2}$ dropping by three orders of magnitude over an atmospheric height separation of $\sim$800~km, confirming that the pore was indeed able to guide wave energy into higher atmospheric layers of the solar atmosphere. Standing slow sausage waves have also been detected in pores by Freij et al. \cite{Freij2016}, who characterised the waves as `standing' due to the estimated vertical wavelengths that would indicate strong reflection at the transition region boundary --- i.e., the creation of a chromospheric resonator that has recently been identified in sunspots \cite{Jess2020}.

More recently, Keys et al. \cite{Keys2018} examined the spatial distributions of wave power contained within solar pores, and were able to provide further classification of these waves in terms of surface- or body-mode sausage waves. The authors made use of seven pore data sets, containing high cadence observations obtained with the ROSA instrument, to demonstrate that surface modes appear to be more prevalent in pores than body modes, particularly in larger pores. Keys et al. \cite{Keys2018} suggest that higher magnetic field strengths, and therefore a larger gradient between the pore and the surrounding quiet Sun in larger pores, may promote surface modes over body modes.

In this study, we utilise state-of-the-art spectropolarimetric data of a unique solar pore grouping to further advance our understanding of MHD waves in the lower solar atmosphere. We employ spectral line bisector methods to obtain height-dependent velocities that allow us to detect the presence of waves, combined with high precision inversion techniques to track the evolution of wave energy with geometric height.

\begin{figure}
\centering\includegraphics[trim=0.5cm 0cm 2.5cm 0cm, clip=true, width=\textwidth]{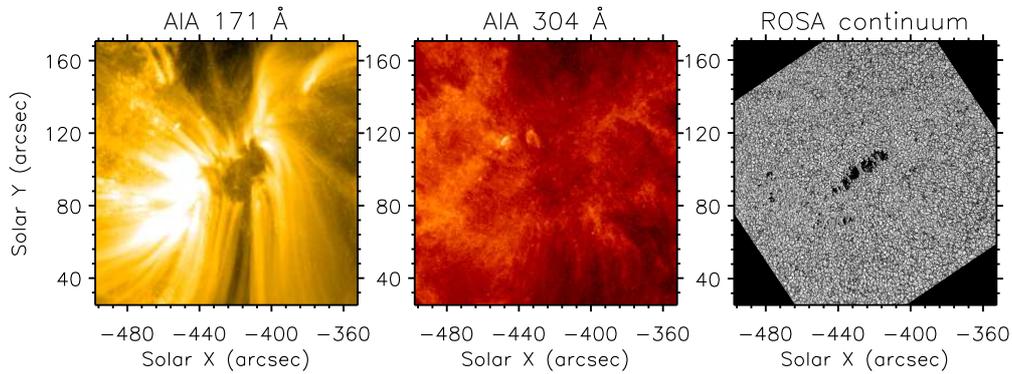}
\caption{AIA/SDO 171{\,}{\AA} (left) and 305{\,}{\AA} (middle) images of active region NOAA~12564, alongside the coaligned ROSA 4170{\,}{\AA} continuum image (right). The axes employed are all solar heliocentric coordinates.}
\label{fig_AIA}
\end{figure}

\section{Observations} \label{section_observations}
The data set presented in this study was acquired between 14:09 -- 15:59~UT on 2016 July 12 with the Facility Infrared Spectropolarimeter (FIRS; \cite{FIRS}), a slit-based spectrograph at the National Science Foundation's Dunn Solar Telescope (DST), Sacramento Peak, New Mexico. The telescope was pointed towards the decaying active region NOAA~12564, positioned at ($-$425$''$, 98$''$) in the solar heliocentric coordinate system, or N10.4E27.5 in the conventional heliographic reference frame. A set of five solar pores were contained within the field of view, positioned along a unique straight-line configuration. To cover all pores in a single exposure, the DST's coude table was rotated so the FIRS slit passed through the centre of each photospheric pore boundary. 

FIRS was employed to obtain high-resolution spectropolarimetric measurements of the photospheric {\SiI} line, which has an associated Land\'{e} factor of $g_{\text{eff}} = 1.5$ \cite{Landeg}.  The slit width corresponded to $0{\,}.{\!\!}{''}225$ on the solar surface, while the spatial sampling along the slit was $0{\,}.{\!\!}{''}15$. The spectral sampling was 0.04{\,}{\AA} per pixel. To increase the polarimetric signal to noise, each time step was the result of the integration of 12 consecutive modulation cycles leading to a 14.6~s cadence. The FIRS data was processed and reduced using the FIRS pipeline provided by the National Solar Observatory (NSO)\footnote{\href{https://nso.edu/telescopes/dunn-solar-telescope/dst-pipelines/}{https://nso.edu/telescopes/dunn-solar-telescope/dst-pipelines/}}. In total, 452~scan steps were obtained across the 110~minute duration of the observing campaign. However, while the seeing conditions remained very good overall, the final 20~minutes of the sequence suffered from periods of seeing degradation. As a result, the data was cropped to the first 90~minutes, providing 355 high-quality spectral scans for subsequent analyses. 

Contextual vector magnetograms and continuum images from the Helioseismic and Magnetic Imager (HMI; \cite{Schou2012}), onboard the Solar Dynamics Observatory (SDO; \cite{Pesnell2012}), were obtained to assist with the co-alignment of our ground-based DST data. The Very Fast Inversion of the Stokes Vector (VFISV; \cite{Borrero2010}) algorithm was applied to SDO/HMI vector data to provide vector magnetogram information to compare with our subsequent FIRS inversions. Conversion of the heliocentric coordinates output by the VFISV algorithm into heliographic projections that are parallel ($B_x$ and $B_y$) and normal ($B_z$) to the solar surface was performed using previously documented techniques \cite{Gary1990}, allowing the true inclination angles of the magnetic fields, with respect to the solar normal, to be uncovered (middle-right panel of Figure~{\ref{fig_FOV}}).

Alongside FIRS, the Rapid Oscillations in the Solar Atmosphere (ROSA; \cite{Jess2010}) instrument was employed to capture simultaneous images of the solar atmosphere through a collection of filters, including the 4170{\,}{\AA} broadband filter used for cross-correlation coalignment with the data products from HMI/SDO. The co-aligned fields of view are displayed in Figure~{\ref{fig_FOV}}. Additional images from the Atmospheric Imaging Assembly (AIA; \cite{Lemen2012}), also onboard the SDO spacecraft, are displayed in Figure~{\ref{fig_AIA}} alongside a ROSA 4170{\,}{\AA} continuum image for comparison. The magnetic connectivity between the photosphere and corona is clearly evident in Figure~{\ref{fig_AIA}}, with the loop structures seen in the coronal 171{\,}{\AA} channel closely linked to the photospheric pore features seen in the ROSA 4170{\,}{\AA} continuum image. However, the AIA/SDO 304{\,}{\AA} transition region filtergram (middle panel of Figure~{\ref{fig_AIA}}) does not contain pronounced pore structures, with the high contrast synonymous with the photospheric images missing in the transition region. This may suggest that the pores can no longer be considered magnetically dominated at chromospheric and transition region heights. As a result, their atmospheres may become plasma dominated and hence the well defined boundaries of the pores may become volume filling, thus losing contrast. This is consistent with the magnetic and plasma pressure interplay as a function of atmospheric height previously discussed in theoretical modelling work \cite{Gary2001}.

\begin{figure}[!t]
\centering\includegraphics[width=0.75\textwidth]{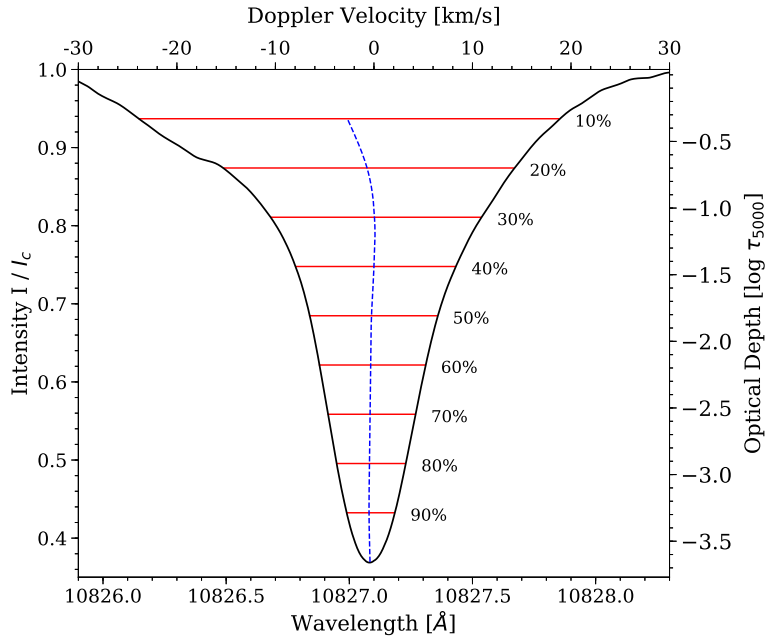}
\caption{An isolated {\SiI} line profile extracted from a quiescent region of the Sun's atmosphere, where the spectral intensity has been normalised by the local continuum intensity, $I_{c}$. Solid red lines indicate line depth percentages, while the dashed blue line traces the bisector wavelength positions at each line depth percentage, which can be converted into Doppler velocities using the scale present at the top of the panel. The scale present at the right-hand side converts the line depth percentages (n.b., {\it{not}} the specific $I/I_{c}$ values) into corresponding optical depths following the conversion suggested by Gonz{\'{a}}lez Manrique et al. \cite{GonzalezManrique}.}
\label{fig_bisectors}
\end{figure}

\section{Analysis} \label{section_analysis}
\subsection{Bisector Velocity Amplitudes}
\label{subsection_bisector}
\noindent The {\SiI} line forms over a large range of atmospheric heights in comparison to most photospheric lines. This allows the line-of-sight (LOS) velocity to be derived at different optical depths (and hence atmospheric layers) using bisector methods \cite{Kulander1966}. Through robust spectral inversions and modelling, recent work has shown that Doppler motions derived at different percentages of the {\SiI} line depth may be coupled to specific optical depths \cite{GonzalezManrique}.

Bisectors of the {\SiI} line profile were computed across 10 -- 90$\%$ of the line depth in intervals of 10\%, with the corresponding wavelength positions, $\lambda$, of each bisector noted. Here, a bisector obtained at `90\%' line depth represents the intensities sampled by the FIRS instrument 90\% of the way towards the darkest part of the line core (i.e., percentage line depth increases as one samples closer to the observed line core). Similarly, a single Gaussian fit was applied to the entire {\SiI} line profile, with the fitted minimum providing the wavelength of the observed line core (i.e., corresponding to 100\% line depth). The measured wavelengths for each component were then transformed into Doppler velocities. This was performed for all slit positions across the full 355 scan steps selected for this study. Figure~{\ref{fig_bisectors}} displays an example {\SiI} line profile, normalised to the continuum intensity, $I_{c}$, with the calculated bisectors depicted using a dashed blue line. Following the work of Gonz{\'{a}}lez Manrique et al. \cite{GonzalezManrique}, the corresponding optical depths are added for clarity, alongside a Doppler velocity conversion for the wavelength shifts measured for each percentage bisector. It can be seen from Figure~\ref{fig_bisectors} that the optical depths span the range $-3.29 < \log(\tau_{\text{500nm}}) < -0.33$ between 10 -- 90\% of the line depth.

\begin{figure}[!t]
\centering\includegraphics[width=0.6\textwidth]{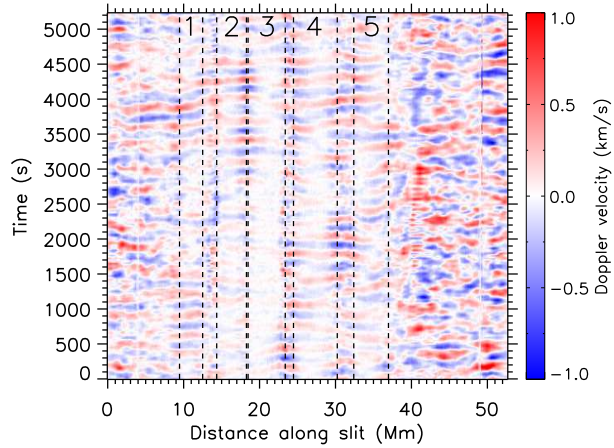}
\caption{Two-dimensional map revealing the evolution of bisector velocities, calculated at 90\% of the {\SiI} line depth, as a function of both distance across the slit and time. Black dashed lines represent the visual photospheric boundaries of each pore, labelled 1--5. Clear wave patterns can be seen throughout by the periodic sign change of the velocity.}
\label{fig_doppler}
\end{figure}

In order to better study the observed Doppler motion, two-dimensional space--time plots at each line depth percentage were created, providing Doppler velocity information as a function of both the distance along the slit and time. Figure~\ref{fig_doppler} displays one such space--time plot for bisector velocities calculated at a line depth of 90\%, corresponding to an optical depth of $\log(\tau_{\text{500nm}}) \approx -3.29$ \cite{GonzalezManrique}. Clear oscillatory motion was detected at all line depths with periods on the order of 5~minutes, which is consistent with the typical $p$-mode spectrum that has previously been demonstrated to be coupled with magnetic structures in the lower solar atmosphere \cite{Lites82, Jess2012}. Oscillations with similar velocity amplitudes, $\sim 0.6$~km/s, have been shown to be related to the presence of slow magnetoacoustic waves that propagate through multiple layers of the solar atmosphere \cite{Prasad2015}. Importantly, as can be seen in Figure~{\ref{fig_doppler}}, the oscillations demonstrate reduced velocity amplitudes within the confines of the pore structures, where the root mean square (RMS) velocities are $<$100~m/s inside the pores, compared with $\sim$300--500~m/s in the surrounding quiet Sun, which are consistent with previous spectroscopic studies \cite{Leighton62, Title92, Felipe2010, Prasad16, Kanoh2016, Oba2017}. To investigate the plasma parameters associated with the detected wave activity, and subsequently evaluate the energy flux, it was imperative to undertake spectropolarimetric inversions of the {\SiI} line.

\subsection{SIR Inversions}
\label{subsection_SIR}

\noindent To probe the wave activity further, atmospheric plasma parameters were inferred through the application of the Stokes Inversion based on Response functions (SIR; \cite{SIR}) code to our {\SiI} Stokes spectra. SIR performs spectropolarimetric inversions under the assumptions of local thermodynamic equilibrium (LTE) and vertical hydrostatic equilibrium, and outputs height-stratified atmospheric parameters that can be related to specific optical depths and/or geometric heights.  The fitting of the Stokes parameters is accomplished by iteratively modifying an initial model atmosphere until the synthesised profiles match the observed spectra. Differences between synthetic and observed profiles are minimised using a Marquart non-linear least-squares algorithm \cite{Numerical_Recipes}. In order to maximise the signal to noise, the {\SiI} spectral data were temporally averaged across a set of time steps during optimal seeing conditions spanning approximately 2~minutes (116.8~s). A time interval of $\approx$2~minutes was chosen since this is long enough to generate high signal-to-noise Stokes signals, yet short enough to prevent solar evolution from contaminating the averaged Stokes profiles. This resulted in a high-quality [$x, \lambda, S$] cube, where $x=355$ is the number of pixels along the length of the slit, $\lambda$ is the wavelength domain, and $S$ represents the four Stokes vectors. This cube, averaged over $\approx2$~minutes during optimal seeing conditions, was then employed in subsequent SIR inversions to provide plasma parameters representative of the background atmosphere.

All pore pixels along the slit were subsequently inverted using an initial atmosphere based on the umbral model for a small sunspot \cite{Hot11Model}. This was considered appropriate since the pore structures under investigation are smaller and have a lower magnetic field strength than a large-scale sunspot, and are therefore likely to have higher background temperatures due to their reduced ability to suppress convective motions. In the surrounding quiescent regions away from the pore structures, the FAL-C model was used for the initial atmosphere \cite{FALCModel}. Stokes $Q$ and $U$ profiles were not considered in the inversions as the pore magnetic fields are predominantly vertical in the lower photosphere (see, e.g., the middle-right panel of Figure~{\ref{fig_FOV}}), resulting in the Stokes $Q$/$U$ signals being relatively weak and prone to fitting difficulties. Three cycles were chosen for the SIR inversion code, with increasing numbers of nodes selected per cycle as shown in Table~{\ref{table_nodes}}. In the first cycle, relatively few nodes were employed in order to gain a solid foundation that future cycles of the code could build upon. Additional nodes were subsequently introduced gradually to allow more freedom for fitting specific line features. For the purpose of investigating the energy flux of the detected waves (see Section~{\ref{subsection_energy_flux}} below), accurate plasma parameters linked to temperature, magnetic field strength, and density are required. 

\begin{figure}[!t]
\centering\includegraphics[width=\textwidth]{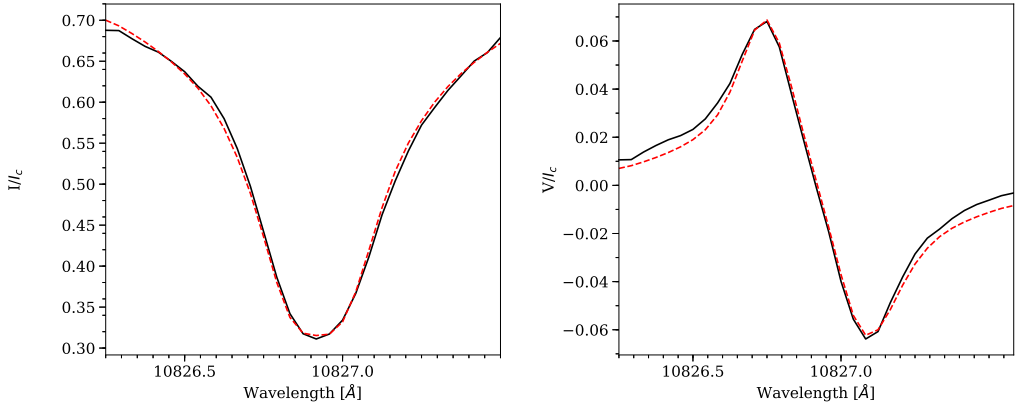}
\caption{Sample {\SiI} Stokes $I$ (left) and $V$ (right) spectra for a pixel inside pore 1, each normalised by the average continuum intensity, $I_{c}$. The black lines show the observed Stokes profiles, while the red dashed lines represent the synthetic profiles generated by SIR. }
\label{fig_stokes}
\end{figure}

Example fits achieved for Stokes $I$ and $V$ profiles for a pixel inside pore 1 are shown in Figure~\ref{fig_stokes}. To evaluate quantitatively the quality of the fits, a reduced $\chi^2$ merit function was calculated following the standard convention~\cite{delToroIniesta2016},
\begin{equation}\label{chi_sq}
    \chi ^2 = \frac{1}{\nu}\sum_{n=0}^{N_{\lambda}} \frac{(I_{syn} - I_{obs})^2}{I_{obs}} \ ,
\end{equation}
where $I_{syn}$ is the intensity of the synthesised Stokes profile at a particular wavelength, $I_{obs}$ is the intensity of the observed Stokes profile at the same wavelength, $N_{\lambda}$ is the number of wavelengths employed in each inversion, and $\nu$ is the degrees of freedom. The average reduced $\chi^2$ values achieved for all inverted pixels were $\overline{\chi}^2 (I/I_{c}) = 0.85$ and $\overline{\chi}^2 (V/I_{c}) = 1.96$, showing excellent fit quality across the field of view.

\begin{table}[!t]
\begin{center}
\caption{Nodes for each atmospheric parameter, alongside Stokes profile weightings used in each inversion cycle of the SIR code.}
\label{table_nodes}
\begin{tabular}{|l||c|c|c|}
\hline
\multicolumn{4}{|c|}{\textbf{Nodes}} \\
\hline
Parameter  &Cycle 1 &Cycle 2 &Cycle 3 \\
\hline
Temperature & 2 & 3 & 4 \\
LOS Velocity & 1 & 2 & 3 \\
Magnetic Field & 1 & 2 & 3 \\
Inclination & 1 & 2 & 3 \\
Microturbulence & 1 & 1 & 1 \\
\hline
\multicolumn{4}{|c|}{\textbf{Weights}} \\
\hline
Stokes $I$ & 1 & 1 & 1 \\
Stokes $V$ & 1 & 1 & 1 \\
\hline
\end{tabular}
\end{center}
\vspace*{-4pt}
\end{table}

SIR computes atmospheric parameters across numerous optical depths, whose corresponding geometric heights are also derived under the assumption of hydrostatic equilibrium. There is some natural uncertainty surrounding the conversion between optical depth and geometric height due to the assumptions that are made, with the reliability of the geometric height scale being reliant on accurate values for temperature, density, and gas pressure. Note that a top boundary condition must be specified for the gas pressure, resulting in the ability to shift the geometric height scale upwards or downwards based on the requirements of hydrostatic equilibrium. Recent research has shown that the precision of gas pressure and density estimations may be improved through the application of a magneto-hydrostatic equilibrium \cite{Borrero2019}, but this is an ongoing area of research and therefore not implemented in the current SIR inversions. 

Hence, it must be noted that varying opacities found for different solar structures, and thus for each pixel along the slit, result in a non-uniform relationship between geometric height and optical depth. To overcome this, each pixel along the slit had its own independent conversion between optical depth and geometric height, which was particularly necessary for the conversion between {\SiI} line depth percentages into optical depths (via the study by Gonz{\'{a}}lez Manrique et al. \cite{GonzalezManrique}), and then subsequently into physical geometric heights. This ensures that the velocity information derived from the line bisectors can be compared directly with the SIR inversion outputs, helping to generate two-dimensional plots of temperature, magnetic field strength, density, and velocity as a function of both distance along the slit and atmospheric height.

Figure~\ref{fig_SIR_params} displays such two-dimensional plots for the magnetic field strength, temperature, plasma density, and {\SiI} RMS velocity, where it can be seen that higher magnetic field strengths and lower temperatures are present in the pores. Importantly, the magnetic field strengths inferred from the SIR code are consistent with those present in the HMI/SDO observations (see the upper-right panel of Figure~{\ref{fig_FOV}}), highlighting the suitability of the time-averaged SIR maps to be used as proxies for the background photospheric solar atmosphere. For the RMS velocity plot (lower-right panel of Figure~{\ref{fig_SIR_params}}), RMS velocities are calculated across the FIRS slit for each of the line depth percentages used in Section~3{\ref{subsection_bisector}}, with the percentage line depths subsequently converted into optical depths and geometric heights as discussed above. The results displayed in Figure~{\ref{fig_SIR_params}} are consistent with previous photospheric studies of magnetic phenomena \cite{Grant2015, Kanoh2016}, but are now displayed as two-dimensional plots to aid visual clarity.

\section{Discussion}
\label{subsection_energy_flux}
\noindent The energy flux, $E$, of a propagating slow magnetoacoustic wave can be calculated following,
\begin{equation}
\label{eq_energy_flux}
E ~=~ \rho~v_{\text{g}}~\langle v^2 \rangle \ ,
\end{equation}
where $\rho$ is the plasma density, $v_{\text{g}}$ is the group velocity of the propagating wave, and $\langle v^2 \rangle$ is the mean square velocity \cite{Bruner1978}. 

\begin{figure}[!t]
\centering
\includegraphics[width=0.49\textwidth]{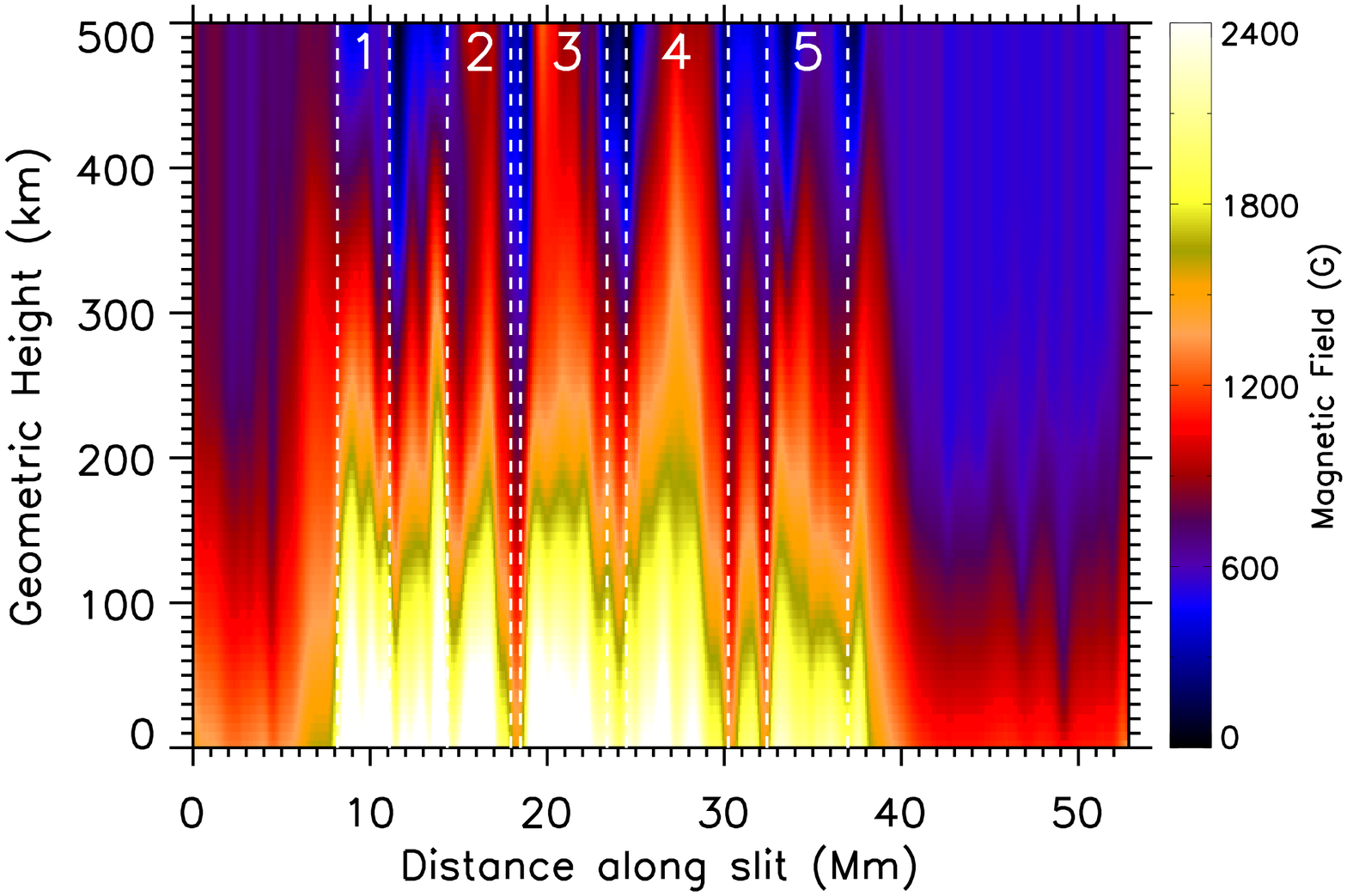}
\includegraphics[width=0.49\textwidth]{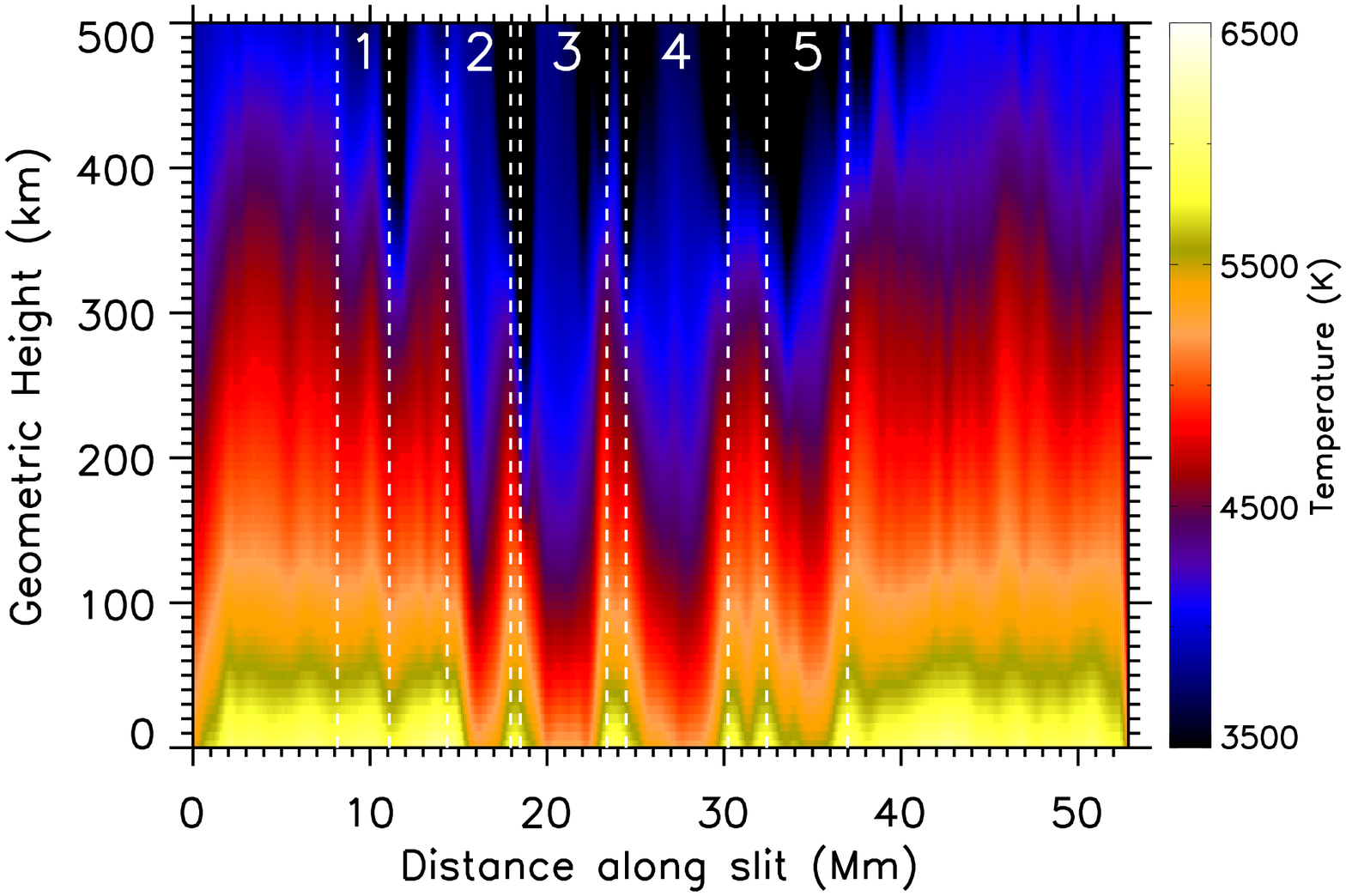}
\includegraphics[width=0.49\textwidth]{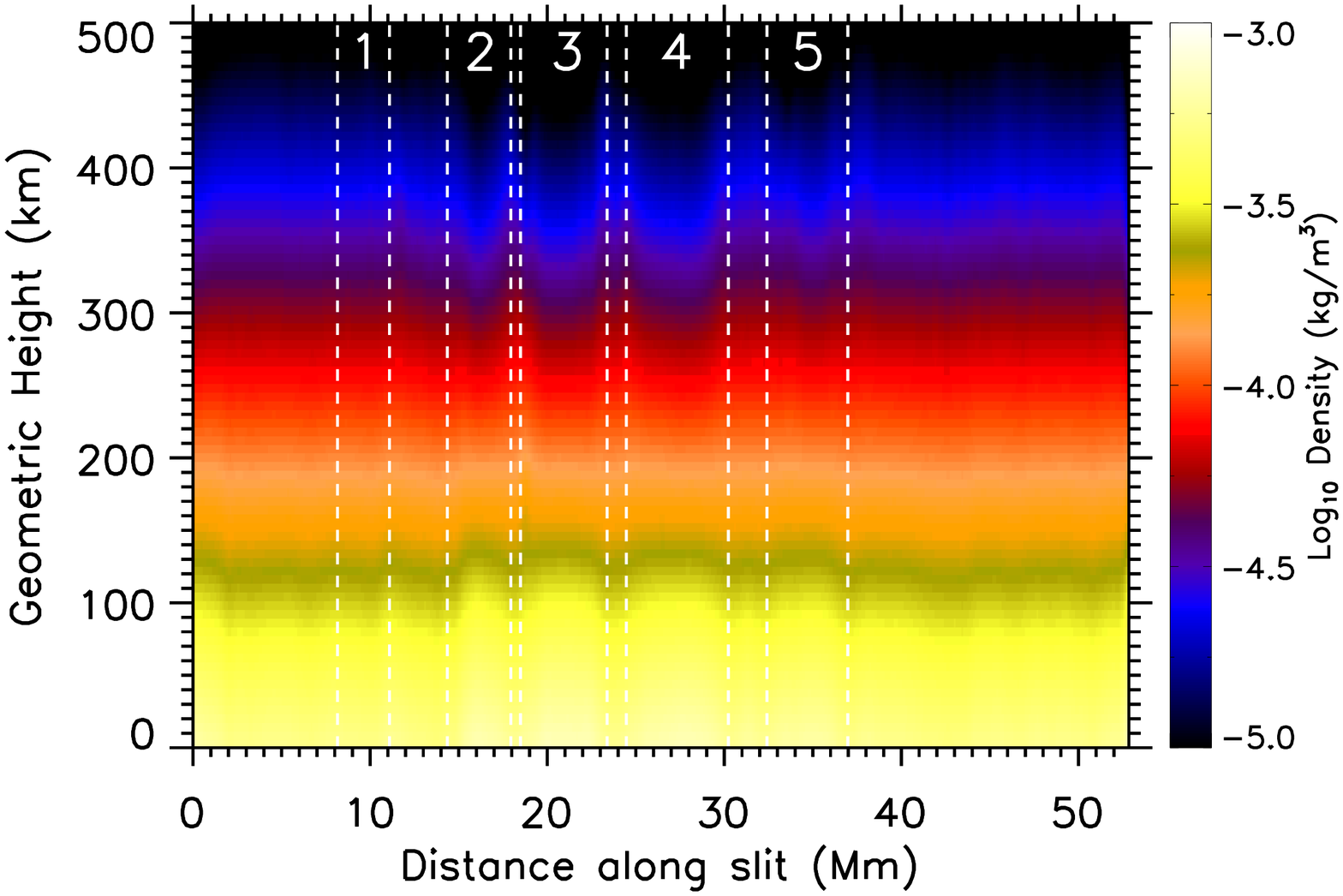}
\includegraphics[width=0.49\textwidth]{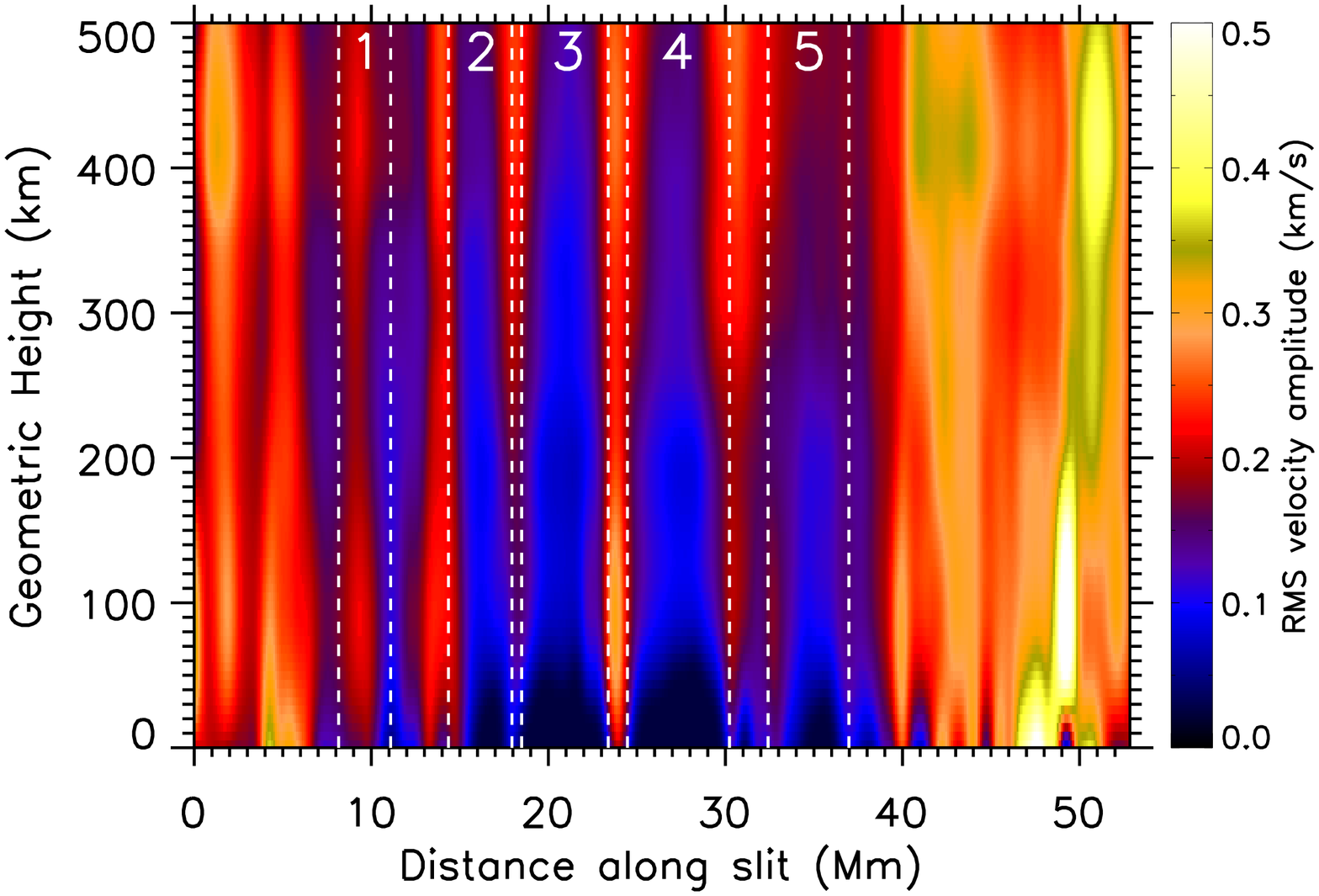}
\caption{Atmospheric parameters as a function of both distance along the FIRS slit and atmospheric height, where the magnetic field strength (upper left), temperature (upper right), and plasma density (lower left) values are computed via SIR inversions of the {\SiI} spectropolarimetric data. The plasma density is displayed on a logarithmic scale to aid visual clarity. The RMS velocity amplitudes (lower right) are calculated from bisector velocities over a range of {\SiI} percentage line depths. Pore boundaries are highlighted by the vertical white dashed lines, with each pore numbered 1--5 as initially defined in the lower panel of Figure~{\ref{fig_FOV}}.} 
\label{fig_SIR_params}
\end{figure}

The plasma density has already been obtained through SIR inversions of the spectropolarimetric {\SiI} data described in Section~3\ref{subsection_SIR}, and is displayed in the lower-left panel of Figure~{\ref{fig_SIR_params}}. Similarly, the mean square velocities can be found by squaring the RMS velocity signals depicted in the lower-right panel of Figure~{\ref{fig_SIR_params}}. 

For a slow magnetoacoustic wave, the group velocity can be characterised by the tube speed. In the thin flux tube approximation \cite{EdwinRoberts83}, the tube speed, $c_T$, is equal to,
\begin{equation}
\label{eqn:tube}
    c_T = \frac{c_s v_A}{\sqrt{c_s^2 + v_A^2}} \ ,
\end{equation}
where $c_s$ and $v_A$ are the local sound and Alfv{\'{e}}n speeds, respectively \cite{Roberts1981, Roberts2005}. The sound speed is given by,
\begin{equation}
    c_s = \sqrt{\frac{\gamma k T}{m}} \ ,
\end{equation}
where $\gamma=5/3$ is the ratio of specific heats, $k$ is the Boltzmann constant, $T$ is the temperature, and $m=2.078 \times 10^{-27}$~kg is the mean photospheric ion mass assuming a composition of 74.9\% hydrogen and 23.8\% helium \cite{Lodders2003}. The Alfv{\'{e}}n speed can be calculated via,
\begin{equation}
\label{eqn:Alfven}
    v_A=\frac{B}{\sqrt{\mu_0 \rho}} \ ,
\end{equation}
where $B$ is the absolute magnetic field strength, and $\mu_0$ is the magnetic permeability.

\begin{figure}[!t]
\centering
\includegraphics[width=0.49\textwidth]{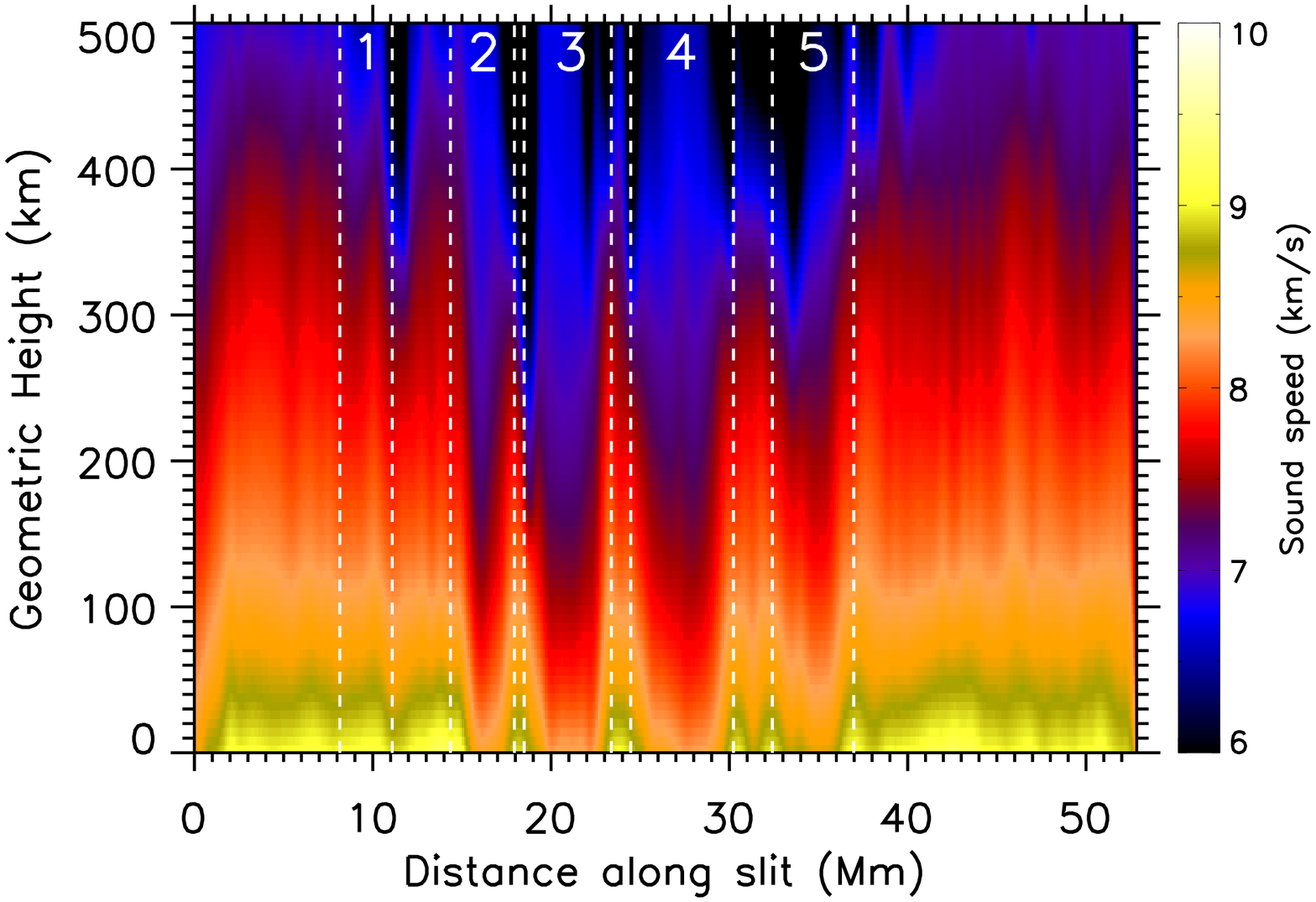}
\includegraphics[width=0.49\textwidth]{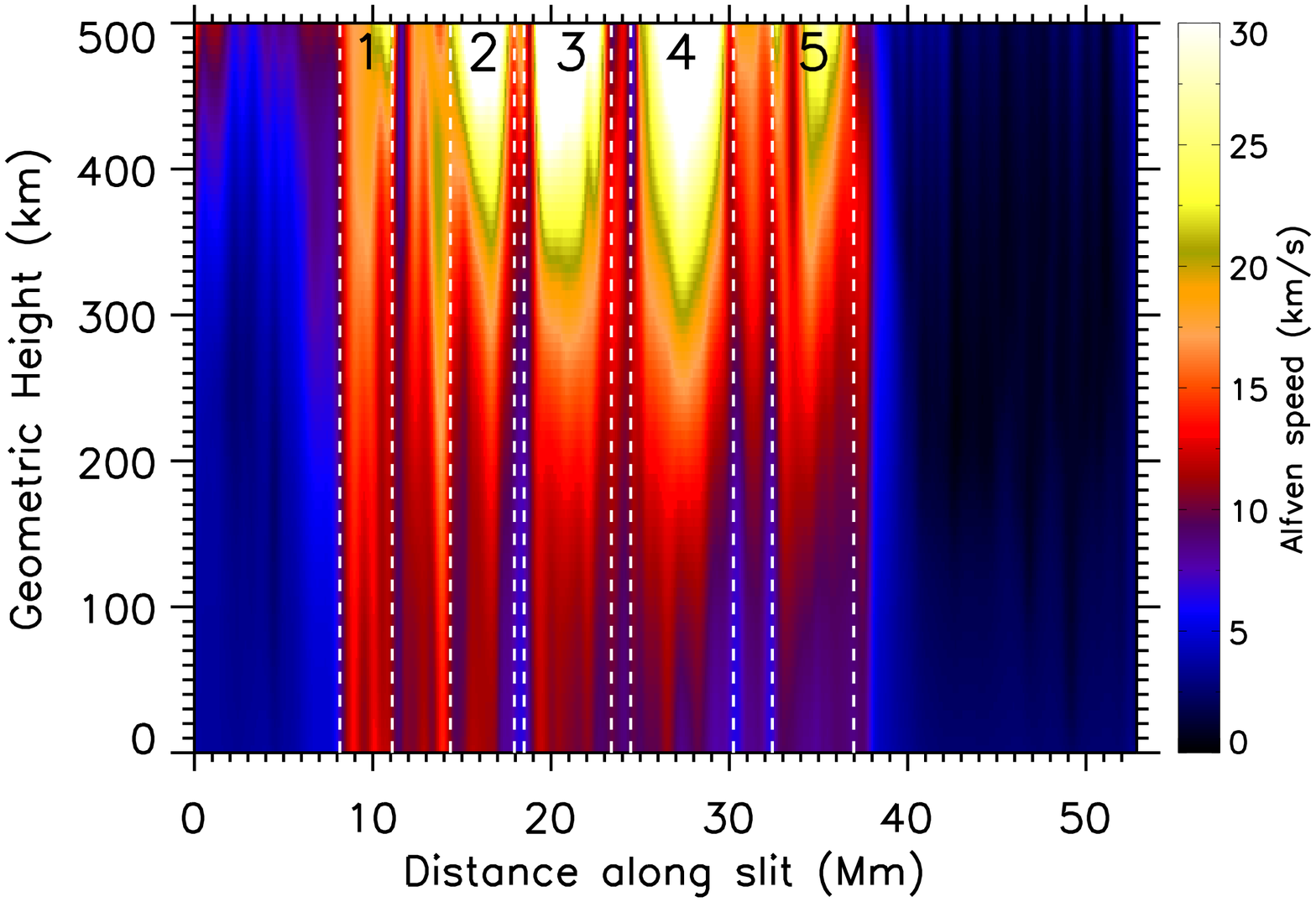}
\includegraphics[width=0.49\textwidth]{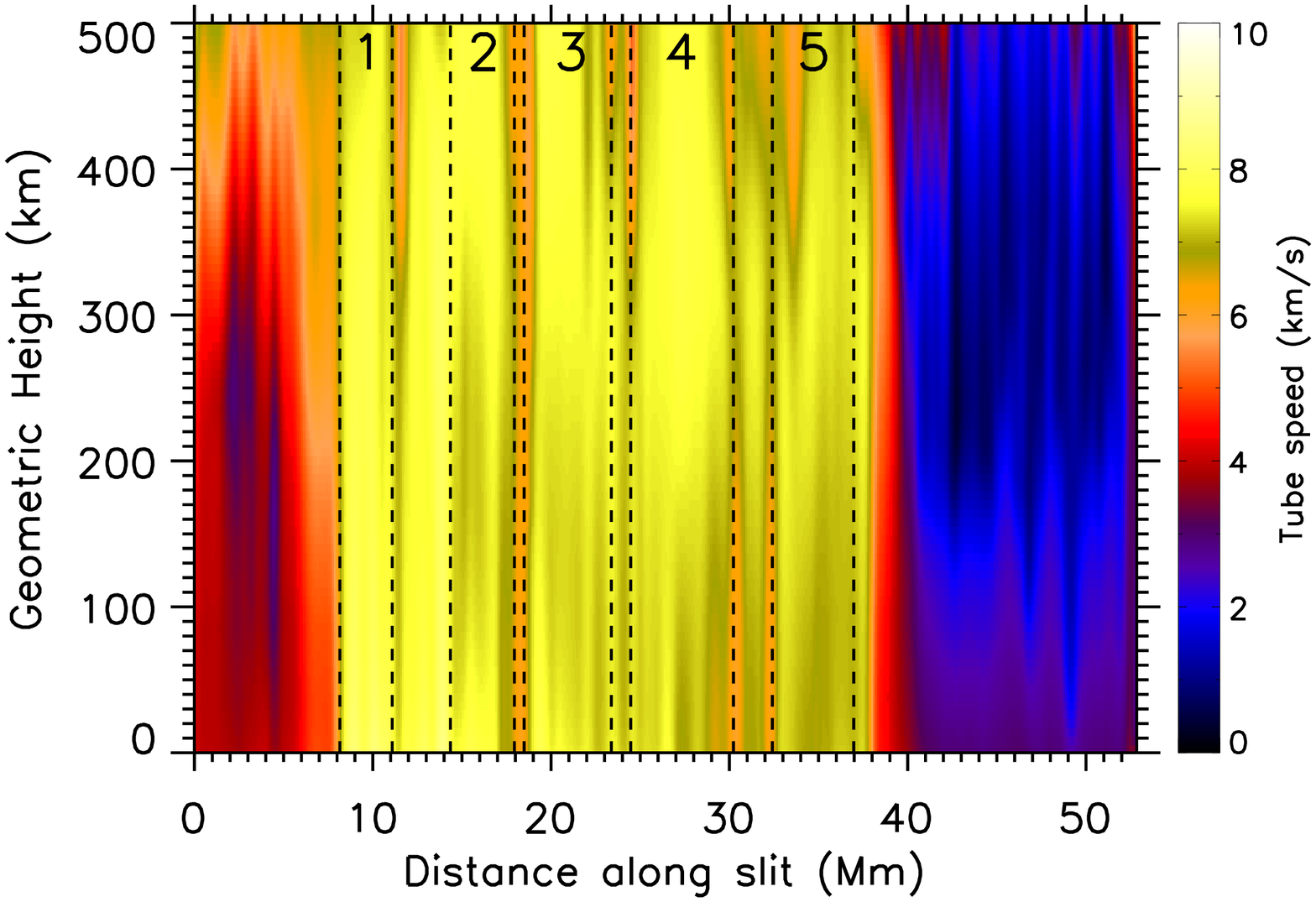}
\caption{The sound (upper left), Alfv\'en (upper right), and tube (bottom) speeds displayed as a function of both distance across the slit and atmospheric height. Pore boundaries are highlighted by the vertical dashed lines, with each pore numbered 1--5 as initially defined in the lower panel of Figure~{\ref{fig_FOV}}.} 
\label{fig_speeds}
\end{figure}

The plasma parameters required to compute the sound, Alfv{\'{e}}n, and tube speeds --- notably the temperatures, densities, and magnetic field strengths --- have already been extracted from the spectropolarimetric SIR inversions and are displayed in Figure~{\ref{fig_SIR_params}}. Following Equations~{\ref{eqn:tube}} -- {\ref{eqn:Alfven}}, the sound, Alfv{\'{e}}n, and tube speeds were computed as a function of both distance along the FIRS slit and atmospheric height, which are displayed in Figure~{\ref{fig_speeds}}. It can be seen in the upper-left panel of Figure~{\ref{fig_speeds}} that the cooler temperatures found inside the pore boundaries result in reduced sound speeds in these locations. Locations away from the pores have sound speeds on the order of 9~km/s, which is consistent with previous numerical modelling efforts of granulation \cite{Jess12}. However, the increased magnetic field strengths found within the pores (upper-right panel of Figure~{\ref{fig_speeds}}) provides increased Alfv{\'{e}}n speeds compared to the quiet Sun surroundings, especially at higher photospheric heights where the density has decreased by multiple orders of magnitude. The Alfv\'en speeds contained within the pores span the range of approximately $5~\text{km/s} \lesssim v_A \lesssim 30~\text{km/s}$, which is consistent with previous pore modelling studies \cite{Grant2015}. Finally, the tube speeds, which are a product of the sound and Alfv{\'{e}}n speeds (see Equation~{\ref{eqn:tube}}), show increased speeds within the pore boundaries across all atmospheric heights, highlighting the ability of magnetic pores to propagate wave energy flux through the atmosphere with elevated group velocities.

\begin{figure}[!t]
    \centering
    \includegraphics[width=0.7\textwidth]{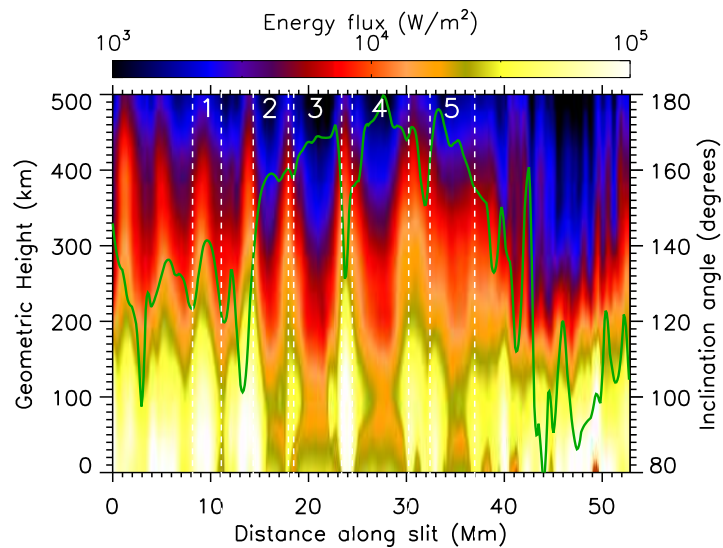}
    \caption{Energy flux, in units of W/m$^2$, displayed as a function of both distance across the slit and atmospheric height. To aid visualisation, the colour scale employed is a logarithmic scale spanning $10^3 - 10^5$~W/m$^2$. Pore boundaries are highlighted by the vertical white dashed lines, with each pore numbered 1--5 as initially defined in the lower panel of Figure~{\ref{fig_FOV}}. The solid green line displays the inclination angles of the magnetic field along the slit, with angles of $90^{\circ}$ and $180^{\circ}$ highlighting fields that are parallel and perpendicular, respectively, to the solar surface. High energy fluxes towards the edges of pore structures (i.e., pores 2, 3, and 4) may be the result of surface mode waves, while more uniform energy structuring (i.e., pores 1 and 5) may be related to the presence of body mode waves. Energy flux values seen in non-pore regions are likely a result of convective overshoots and not specifically propagating magnetoacoustic wave phenomena.}
    \label{fig_energy_flux}
\end{figure}

With the tube speeds computed as a function of both distance along the slit and atmospheric height (lower panel of Figure~{\ref{fig_speeds}}), it is now possible to combine them with the plasma density and RMS velocity maps shown in the lower-left and lower-right panels of Figure~{\ref{fig_SIR_params}}, respectively, to calculate the energy flux following Equation~{\ref{eq_energy_flux}}. The corresponding energy flux map is displayed in Figure~\ref{fig_energy_flux}. Evidence of wave energy flux is visible both within the pore regions labelled 1--5, in addition to the more quiescent regions of the Sun outside of the exterior pore boundaries. The regions outside the pore boundaries are still magnetic in nature, with $B \neq 0$~G (upper-left panel of Figure~{\ref{fig_SIR_params}}), resulting in positive tube speeds (i.e., $c_{T} > 0$~km/s) everywhere along the FIRS slit (lower panel of Figure \ref{fig_speeds}). This results in wave energy flux, through the application of Equation~{\ref{eq_energy_flux}}, being present throughout the entire observed atmosphere. However, the calculated energy fluxes in these locations may not entirely be related to guided magnetoacoustic wave motion. The larger RMS velocities in the non-pore regions (see the lower-right panel of Figure~{\ref{fig_SIR_params}}) may be a result of other non-oscillatory dynamics, such as convective overshoots \cite{Wedemeyer08, Schmitt84}, which may manifest as artificially heightened energy flux values in Figure~{\ref{fig_energy_flux}}. The higher energy fluxes calculated in the quieter solar region to the south-east of the pores ($<$7~Mm), when compared to the quiet Sun region to the north-west ($>$43~Mm), is mostly likely a result of the stronger total magnetic field strengths found in this location (see the upper-right panel of Figure~{\ref{fig_FOV}} and the upper-left panel of Figure~{\ref{fig_SIR_params}}). A higher magnetic field strength naturally results in increased Alfv\'en and tube speeds for that region. This is confirmed in Figure~{\ref{fig_speeds}}, where increased Alfv\'en and tube speeds are preferentially present in the south-east ($<$7~Mm) region of the field of view. As a result, larger tube speeds, coupled with similar densities and velocity fluctuations, will manifest as increased energy fluxes, even if they are not strictly coupled to upwardly propagating magnetoacoustic wave phenomena. In these locations (e.g., $>$38~Mm in Figure~{\ref{fig_energy_flux}}) we see a much more rapid decrease in energy flux with atmospheric height when compared with the pore regions. This may be a direct consequence of the more heavily inclined magnetic fields present in these locations, hence inhibiting the propagation of energy flux (created through wave motion and/or convective overshoots) into higher regions of the solar atmosphere.  

The inclination angles of the magnetic field, which are derived from the HMI/SDO VFISV vector magnetograms and correspond to a height range of $250-300$~km \cite{Norton2006, Fleck2011}, are displayed in Figure~{\ref{fig_energy_flux}} using a solid green line. Inclination angles, which were calculated from the heliographic re-projections of the HMI/SDO vector magnetograms, were preferred here due to these inclinations being relative to the solar normal, unlike the inclination angles output from SIR inversions which are relative to the line-of-sight of the observer. It can be seen that the pore structures have magnetic fields approximately normal to the solar surface (consistent with the HMI/SDO observations shown in the middle-right panel of Figure~{\ref{fig_FOV}}), while locations away from the pores demonstrate more heavily inclined magnetic fields, with regions $>43$~Mm along the FIRS slit having approximately horizontal magnetic field configurations. This highlights the efficiency of magnetic pores as wave energy conduits in the lower solar atmosphere.

Another notable feature visible in Figure~{\ref{fig_energy_flux}} is the increase in energy flux visible towards the edges of the pore boundaries for pores 2, 3, and 4. Contrarily, it can be seen in Figure~{\ref{fig_energy_flux}} that pore 1 has the dominant energy flux towards its centre, while pore 5 displays more uniform energy structuring across its entire diameter. 

Such energy structuring can be attributed to the presence of surface (pores 2, 3, and 4) and body (pores 1 and 5) mode waves, which have only recently been conclusively observed in photospheric imaging data \cite{Keys2018}. Previous identifications of surface mode waves have come from theoretical estimations \cite{Grant2015, Moreels15b, Yu2017, Chen2018}. However, here we clearly show the existence of heightened wave energy towards the edges of the pore boundaries, which has only been made possible by the high spectral, spatial, and temporal resolutions of our dataset coupled with modern spectropolarimetric inversion routines. 

\begin{figure}[!t]
\centering\includegraphics[width=\textwidth]{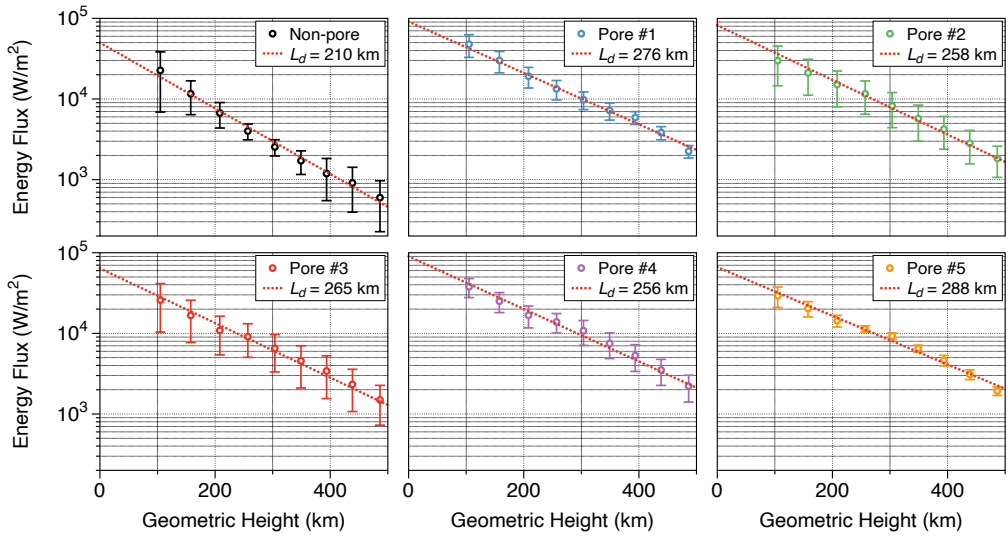}
\caption{Average energy fluxes as a function of geometric height for a non-pore region (upper-left), in addition to pores 1--5 (upper-middle through to lower-right panels). Uncertainties at each geometric height correspond to the standard deviations of the energy fluxes included in the corresponding average. The dotted red lines correspond to exponential lines of best fit, with the corresponding damping lengths, $L_d$, indicated in the legend provided at the top-right of each panel. It can be noted that pores (1 \& 5) containing body mode waves exhibit the longest damping lengths, while pores (2--4) exhibiting surface mode waves demonstrate the shortest damping lengths.}
\label{fig_damping_length}
\end{figure}

A natural question arises as to why pores 1 and 5 demonstrate different wave modes (`body' vs `surface') when compared to the more central pores 2 -- 4. From inspection of the structuring of the pores in Figure~{\ref{fig_FOV}}, the more complex physical composition of pores 1 and 5 may help provide an answer. Pores 1 and 5 are smaller in area than pores 2 -- 4, and display weaker magnetic field strengths at higher atmospheric heights (see the upper-left panel of Figure~{\ref{fig_SIR_params}}). Reduced field strengths will result in less inhibition of convection, and the subsequent higher temperatures in pores 1 and 5 (see the upper-right panel of Figure~{\ref{fig_SIR_params}}) may contribute to the structuring seen in the energy flux map shown in Figure~{\ref{fig_energy_flux}}. For example, the increased temperatures found in pores 1 and 5 result in more elevated sound speeds, which helps negate the effects of reduced Alfv{\'{e}}n speeds on the corresponding tube speeds, where lower Alfv{\'{e}}n speeds are caused by the weaker magnetic field strengths present in these pores (see Figure~{\ref{fig_speeds}}). Hence, the less uniform structuring of pores 1 and 5 may actually contribute to the more even stratification of wave energy flux visible in Figure~{\ref{fig_energy_flux}. 

Indeed, the increased RMS velocity inside pore 1 (see Figure~{\ref{fig_SIR_params}}), which may be related to the presence of small-scale convective motions not suppressed by the weaker magnetic fields, may promote elevated wave energy fluxes towards the centre of these bodies. However, the clear periodicities observed in all pores (see Figure~{\ref{fig_bisectors}}) show unequivocal evidence of upwardly propagating wave motion that could not be replicated simply by the presence of bright structures or small-scale granulation within the confines of the pore. This, alongside the very confined spread of energy fluxes across the diameters of pores 1 and 5 (Figure~{\ref{fig_damping_length}}) would suggest the most likely explanation is that body modes are present in these structures. As a result, we currently do not have an answer as to why pores 1 and 5 demonstrate different wave modes when compared to their more central pore counterparts. Future work will need to examine the sub-photospheric drivers responsible for the observed wave motion, and through intensive numerical modelling (e.g., \cite{Riedl2019}), uncover the role of waveguide structuring, composition, and inclination in the visible wave signatures.}

The energy fluxes (including both body and surface modes), visible in Figure~{\ref{fig_energy_flux}}, demonstrate the efficiency of magnetic pores as magnetoacoustic wave energy conduits in the lower solar atmosphere. Indeed, the energy fluxes measured here as a function of atmospheric height also agree with the findings of Grant et al. \cite{Grant2015}, who employed imaging observations and a theoretical framework to estimate the associated energies. Previous work has examined the propagation of magnetoacoustic waves in coronal fan loops, and estimated the damping lengths of the wave activity by examining the amplitude decay rates as a function of distance along the waveguide \cite{DeMoortel2003, DeMoortel2004, DeMoortel2004b, Marsh2011, KrishnaPrasad2012, KrishnaPrasad2014}. Specifically, the relation used to uncover the damping length, $L_d$, is given by \cite{KrishnaPrasad2019},
\begin{equation}
    A(h) = A_0 e^{-h/L_d} \ ,
\end{equation}
where $A(h)$ is the amplitude measured at a given height, $h$, and $A_0$ is the initial amplitude of the wave. As the energy of the propagating magnetoacoustic waves is proportional to the amplitude squared, i.e., $E(h) \propto A(h)^2$, we can adapt the damping length equation to be compatible with the energy fluxes displayed in Figure~{\ref{fig_energy_flux}}, where, 
\begin{equation}
\label{eqn:dampinglength}
    E(h) = E_0 e^{-2h/L_d} \ .
\end{equation}

For the first time, by employing Equation~{\ref{eqn:dampinglength}}, we are able to calculate the photospheric damping lengths of propagating magnetoacoustic waves in solar pores. Figure~{\ref{fig_damping_length}} displays the average energy fluxes for each of the 5 pores, alongside a quiescent region away from the pore structures (averaged over $43 - 50$~Mm along the FIRS slit where the magnetic field inclinations are less vertical; solid green line in Figure~{\ref{fig_energy_flux}}), as a function of atmospheric height. The average energy flux for each pore was calculated by taking the mean value at each geometric height between the vertical dashed lines that outline the respective pore boundaries in Figures~{\ref{fig_SIR_params}} -- {\ref{fig_energy_flux}}. For each atmospheric height, the energy uncertainties in Figure~{\ref{fig_damping_length}} correspond to the standard deviations of the averaged energies. It can be seen that pores 1 and 5, which are best characterised as containing body mode waves, have the smallest energy uncertainties as a result of the relatively uniform energy structuring spanning their entire cross-section. Importantly, our energy fluxes also corroborate the work of Keys et al. \cite{Keys2018}, who suggested that surface mode sausage waves should carry more energy flux when compared to their body mode counterparts. Lines of best fit allow the damping lengths to be calculated for each of the structures studied. The quiescent region, located away from the pores, displays the most rapid damping length with $L_d \approx 210$~km. This may be a consequence of the reduced magnetic field strengths (and hence tube speeds) and larger inclination angles (see the middle-right panel of Figure~{\ref{fig_FOV}}) being unable to efficiently guide the magnetoacoustic waves upwards. Pores 1 and 5, which are best characterised as containing body mode waves, have the longest damping lengths, corresponding to $L_d \approx 276$~km and $L_d \approx 288$~km, respectively. On the other hand, pores 2, 3, and 4, which are best characterised as surface mode waveguides, have slightly smaller damping lengths equating to $L_d \approx 258$~km, $L_d \approx 265$~km, and $L_d \approx 256$~km, respectively. Recent theoretical work has shown that the degree of magnetic twist in the waveguide can drastically affect the damping rate of surface sausage mode waves \cite{Sadeghi2019}. Furthermore, the interplay between electric resistivity and resonant absorption may also contribute to the differing damping lengths found between the surface- and body-mode waves \cite{Chen2018}.

The damping lengths calculated here are much shorter than those deduced for the corona (approximately $3700 - 4800$~km; \cite{KrishnaPrasad2019}), which is likely a consequence of the much shorter density and pressure scale heights found in the lower solar atmosphere \cite{Stein1989}. Following the standard convention \cite{Mattig1969} for the scale height, $H$, where,
\begin{equation}
    H = \frac{\Delta h}{\Delta \log (\tau_{\text{500nm}})} \ ,
\end{equation}
the pores in the present study span a geometric height of $\Delta h = 500$~km and an optical depth of $\Delta \log (\tau_{\text{500nm}}) \approx 2.96$, giving $H \approx 170$~km. As a result, the average damping length across all 5 pores, $\overline{L_d} \approx 268$~km, is on the same order of magnitude as the estimated photospheric scale height. Modelling work has predicted coronal scale heights on the order of 30 -- 60~Mm ($\sim$30{\,}000 -- 50{\,}000~km; \cite{Aschwanden2001, Singh2002, Alissandrakis2019}), so an initial comparison would suggest that the measured coronal damping lengths are much shorter than typical coronal scale heights (by at least an order of magnitude). However, recent theoretical work has calculated the expected coronal damping lengths using classical Spitzer values for thermal conduction and predicted much longer damping lengths than observed in coronal loops \cite{KrishnaPrasad2019}, hence suggesting either a lack of sensitivity in current coronal observations, or ill-constrained values of thermal conduction in current modelling efforts. Regardless, this is an area of active research, and our measurements of the damping lengths in the lower solar atmosphere will be crucial for comparisons to the next generation of numerical models of MHD wave activity. 

\section{Concluding Remarks}

Here we have presented spectropolarimetric {\SiI} data of a collection of five magnetic pores acquired by the FIRS instrument at the DST. Our observations show clear evidence of wave activity occurring in the lower solar atmosphere, with a bisector velocity study of the {\SiI} line providing velocity oscillations visible across a variety of optical depths with periods on the order of 5~minutes, consistent with the global $p$-mode spectrum. Through spectropolarimetric inversions, we deduce the local plasma densities, temperatures, and magnetic field strengths as a function of atmospheric height, which allowed us to calculate the corresponding sound, Alfv\'en, and tube speeds across our field of view. Parameters derived from the inversions, combined with root mean square velocities, provides energy flux estimates as a function of height in the lower solar atmosphere. We find evidence of considerable energy flux trapped within the pore boundaries, with on the order of 30~kW/m$^{2}$ at a geometric height of $\approx$100~km. Examination of the energy structuring across the diameter of each pore reveals evidence for both propagating surface- and body-mode sausage waves, something that supports recent imaging observations \cite{Keys2018} with spectropolarimetry. 

For the first time, we estimate the damping lengths of the propagating sausage mode waves trapped within the five pores. We find that the pores demonstrating surface mode characteristics have an average damping length of $\overline{L_d} \approx 260$~km, while the pores identified as displaying body mode signatures have a slightly longer average damping length of $\overline{L_d} \approx 282$~km. Both of these values are on the same order of magnitude as the photospheric scale height, which we estimate to be $H \approx 170$~km. We hypothesise that the surface mode waves may be able to dissipate their energy more readily, through a combination of magnetic twist, electric resistivity and/or resonant absorption mechanisms \cite{Sadeghi2019, Chen2018}. We appreciate that the measured difference in damping lengths is small and based around a limited number of pores. As a result, we openly encourage the community to continue this type of study on a larger, more statistically significant basis. It is hoped that new high-resolution facilities, such as the Daniel K. Inouye Solar Telescope (DKIST; \cite{Tritschler2016}), with improved polarimetric sensitivity will allow more precise measurements to be made. For example, the Diffraction Limited Near Infrared Spectropolarimeter (DL-NIRSP) instrument will enable simultaneous two-dimensional spatial mapping of the {\SiI} line to be obtained with a spatial resolution as high as $0{\,}.{\!\!}{''}06$ (some five times higher than the data presented in our current study). The greater spatial resolution will enable the precise energy structuring of the pores to be deduced with greater clarity, and the two-dimensional fibre-fed nature of the instrument will allow the entire pore structure to be studied simultaneously. As a result, we look forward to upcoming DKIST observations to shine light on the propagation of sausage mode waves in solar pores, and uncover their true role in supplying energy to the outer solar atmosphere.

\enlargethispage{20pt}


\dataccess{The data used in this paper are from the observing campaign entitled {\it{`The Influence of Magnetism on Solar and Stellar Atmospheric Dynamics'}} (NSO-SP proposal T1081; principal investigator: DBJ), which employed the ground-based Dunn Solar Telescope, USA, during July 2016. The Dunn Solar Telescope at Sacramento Peak/NM was operated by the National Solar Observatory (NSO). NSO is operated by the Association of Universities for Research in Astronomy (AURA), Inc., under cooperative agreement with the National Science Foundation (NSF). Additional supporting observations were obtained from the publicly available NASA’s Solar Dynamics Observatory (\href{https://sdo.gsfc.nasa.gov}{https://sdo.gsfc.nasa.gov}) data archive, which can be accessed via \href{http://jsoc.stanford.edu/ajax/lookdata.html}{http://jsoc.stanford.edu/ajax/lookdata.html}. The data that support the plots within this paper and other findings of this study are available from the corresponding author upon reasonable request.}

\aucontribute{DBJ carried out the experiments and conceived of and designed the study. CAG-M performed the data reduction and scientific analysis, with assistance from DBJ, SDTG, PHK, CB, SJ, and BRC. CAG-M drafted the manuscript, with theoretical input provided by JMR and TVD. All authors read and approved the manuscript.}

\competing{The authors declare that they have no competing interests.}

\funding{Funding has come from the following sources: 
\vspace{-3mm}
\begin{itemize}
    \item Invest NI and Randox Laboratories Ltd. Research \& Development Grant (059RDEN-1);
    \item European Union's Horizon 2020 research and innovation programme (grant agreement no. 682462);
    \item Research Council of Norway through its Centres of Excellence scheme (project no. 262622);
    \item European Union's Horizon 2020 research and innovation programme (grant agreement No 724326);
    \item Research Council of Norway (project number 262622); and
    \item The Royal Society (grant Hooke18b/SCTM).
\end{itemize}}

\vspace{-3mm}
\ack{CAG-M, DBJ, and SDTG are grateful to Invest NI and Randox Laboratories Ltd. for the award of a Research \& Development Grant (059RDEN-1) that allowed the computational techniques employed to be developed. SJ acknowledges support from the European Research Council under the European Union's Horizon 2020 research and innovation programme (grant agreement no. 682462) and from the Research Council of Norway through its Centres of Excellence scheme (project no. 262622). JMR and TVD were supported by the European Research Council (ERC) under the European Union's Horizon 2020 research and innovation programme (grant agreement No 724326). TVD received further support from the C1 grant TRACEspace
of Internal Funds KU Leuven (number C14/19/089). The space-based data employed in this work is courtesy of NASA/SDO and the AIA, EVE, and HMI science teams. The authors wish to acknowledge scientific discussions with the Waves in the Lower Solar Atmosphere (WaLSA; \href{www.WaLSA.team}{www.WaLSA.team}) team, which is supported by the Research Council of Norway (project number 262622), and The Royal Society through the award of funding to host the Theo Murphy Discussion Meeting {\it{``High resolution wave dynamics in the lower solar atmosphere''}} (grant Hooke18b/SCTM).}



\end{document}